# GAMMA-RAY BURST LIGHT CURVE RECONSTRUCTION WITH PREDICTIVE MODELS

Zhunuskanov A., Sakan A.*, Akhmetali A., Zaidyn M., Ussipov N.

Department of Electronics and Astrophysics, Al-Farabi Kazakh National University, Almaty, Kazakhstan
**Corresponding author*: aknursakan47@gmail.com

***Abstract.*** *Gamma-ray bursts represent some of the most energetic and complex phenomena in the universe, characterized by highly variable light curves that often contain observational gaps. Reconstructing these light curves is essential for gaining deeper insight into the physical processes driving such events. This study proposes a machine learning-based framework for the reconstruction of gamma-ray burst light curves, focusing specifically on the plateau phase observed in X-ray data. The analysis compares the performance of three sequential modeling approaches: a bidirectional recurrent neural network, a gated recurrent architecture, and a convolutional model designed for temporal data. The findings of this study indicate that the Bidirectional Gated Recurrent Unit model showed the best predictive accuracy among the evaluated models across all gamma-ray burst types, as measured by Mean Absolute Error, Root Mean Square Error, and Coefficient of Determination. Notably, Bidirectional Gated Recurrent Unit exhibited enhanced capability in modeling both gradual plateau phases and abrupt transient features, including flares and breaks, particularly in complex light-curve scenarios.*

## 1. Introduction

Gamma-ray bursts (GRBs) are among the most energetic and transient phenomena in the Universe, characterized by brief flashes of high-energy photons, typically peaking above ~0.1 MeV [1]. These events exhibit a remarkable diversity in duration, ranging from milliseconds to several minutes, as well as highly variable temporal and spectral structures, which pose significant challenges for theoretical modeling. Over the past decade, the field of GRB research has undergone a rapid transformation, evolving from a specialized area of high-energy astrophysics into a major domain of observational cosmology and astrophysics. This progress has been driven by a succession of space missions that enabled both accurate localization and extensive multi-wavelength follow-up of GRBs.

The Burst and Transient Source Experiment (BATSE) onboard the Compton Gamma Ray Observatory provided strong evidence for the isotropic sky distribution of GRBs, implying an extragalactic origin [2]. The Swift Burst Alert Telescope (BAT) [3] detects prompt emission, and the follow-up afterglow is detected using the X-ray [4]. The subsequent BeppoSAX mission enabled the first detections of X-ray afterglows and accurate source localization, which led to the discovery of long-wavelength counterparts and host galaxies [5]. These breakthroughs were further advanced by the High Energy Transient Explorer (HETE-II) [6] and complemented by extensive ground-based observations in the optical, infrared, and radio bands.

The detection of long-lived afterglows has enabled in-depth studies of the circumburst environment and provided compelling evidence for collimated relativistic outflows. Given their extreme luminosities and detectability at high redshifts [8–10], GRBs serve as powerful probes of the distant Universe [11-13]. They offer a unique opportunity to investigate key aspects of cosmology, including the expansion history and the nature of dark energy, the cosmic star formation rate, the timing and processes of reionization, as well as the chemical enrichment of the interstellar and intergalactic media over cosmic time.

Consequently, GRBs are not only key astrophysical phenomena but also valuable tools for probing the Universe on cosmological scales. To fully utilize their potential, precise modeling of their light curves (LCs) is essential. In this context, machine learning (ML) has become an increasingly important tool in astrophysics [14-16], with successful applications in different areas of astronomy and other fields [17-22]. These advancements have led to a growing interest in using ML for GRB LCs reconstruction.

The t-Distributed Stochastic Neighbor Embedding (t-SNE) algorithm has demonstrated stable clustering of GRBs even with partial data removal, highlighting its robustness for dynamic or incomplete datasets [23]. ML has also proven effective in localizing previously unidentified GRBs, revealing uniform sky distributions and suggesting potential magnetar associations [24]. Clustering analyses of Fermi GRBs using spectral features have identified four consistent classes with distinct α and Epeak characteristics, enhancing our understanding of GRB subtypes [25]. Optimization-based modeling has successfully reproduced GRB-SN light curves, supporting magnetar-driven scenarios with physically plausible parameters [26]. Automatic Gaussian Mixture Model (AutoGMM) applied to t-SNE embeddings further revealed coherent density structures, validating the use of unsupervised clustering in GRB mapping [27]. Simulated GRB LCs generated with ML techniques have been shown to statistically match real data, confirming the reliability of data-driven models [28]. Large-scale classification using Uniform Manifold Approximation and Projection (UMAP), t-SNE, and K-means consistently separates Fermi GRBs into two main groups across different parameter spaces [29]. A similar two-cluster structure was observed in a study linking GRB groupings to physical origins, such as compact mergers and collapsars, rather than traditional duration-based classification [30]. These studies collectively demonstrate that ML not only enhances GRB classification but also offers a promising pathway for reconstructing their complex LC.

Recently, Dainotti M. G. et al. [31] applied stochastic reconstruction to GRB LC using Willingale (W07) and broken power-law (BPL) models, along with Gaussian processes. At 10% noise, uncertainties in key parameters-plateau end time, flux, and post-plateau decay index were reduced by up to 43.9%, enhancing the accuracy of GRB-based cosmological studies. Sourav S. et. al. [9] applied a novel approach using bidirectional LSTM proposed for GRB LC reconstruction to address gaps in observational data. Compared to traditional methods (W07, BPL, and their Gaussian Process variants), the Bidirectional Long Short-Term Memory (BiLSTM) model generally produced smoother reconstructions but showed a smaller decrease in flux uncertainty.

R. Falco et. al. [32] presents a quantitative analysis of GRB LCs performed using Principal Component Analysis (PCA) and t-SNE for dimensionality reduction and visualization. Synthetic LC were generated with properties closely matching real ones, showing good overlap in the embedded space. The similarity was further confirmed by statistical comparisons, with L2-norm and Wasserstein distances between histograms of real and synthetic LCs equal to 5.3 and 0.2, respectively. Building upon these recent advancements, the present work introduces a ML-based framework for the reconstruction of GRB LC, focusing on the plateau phase observed in X-ray data. Our approach explores and compares the performance of three distinct architectures: BiLSTM, Bidirectional Gated Recurrent Unit (BiGRU), and Temporal Convolutional Network (TCN). These models are well-suited for sequential data with missing values, enabling more robust reconstruction of LC and improved capture of temporal patterns.

The remainder of this paper is organized as follows: Section 2 describes the data preparation and preprocessing steps; Section 3 outlines the architecture and training process for each ML model; Section 4 presents the evaluation of metrics and experimental results; and Section 5 concludes with a summary.

## 2. Methodology

### *2.1 Data Collection and Preprocessing*

The dataset used in this study consists of GRB LC obtained from the publicly available Swift XRT archive. Each LC was extracted from raw Flexible Image Transport System (FITS) files by extracting key observational parameters, including the observation time, flux, and flux uncertainties. To ensure consistency across the dataset, non-informative or incomplete entries were excluded during the initial data cleaning phase. To restore the temporal order of events, all LC were sorted in ascending order of observation time.

.

Subsequently, time values were normalized by dividing by $10^2$, a transformation that standardizes temporal scales and facilitates stable training across models.

Given the wide dynamic range of GRB flux values, both the flux and its associated uncertainty were transformed into logarithmic space using the formula:

$$F(x) = ln\left(\frac{x}{a}\right) + 1, \quad (1)$$

where (x) - is the flux value and (a) - denotes the minimum non-zero flux in the given sequence. This transformation improves the numerical stability of the learning process by compressing large magnitudes and preserving fine variations in small-scale signals - a critical requirement for deep neural architectures sensitive to input distributions.

To address the inherent sparsity and temporal irregularity of GRB observations, a dense interpolation strategy was applied. Specifically, 19 equally spaced points were interpolated between each pair of consecutive observations. This approach significantly increases the temporal resolution of each sequence and provides more informative context for time-dependent models, such as RNNs and TCNs. The resulting sequences were segmented into fixed-size batches, ensuring uniform input dimensions for training. To prevent information loss due to short sequences, each batch was upsampled via repetition until the required length was met. This strategy enables the model to learn consistent temporal patterns across samples and mitigates overfitting, particularly in the case of limited observational windows. The interpolation strategy, normalization by dividing time by $10^2$, and logarithmic transformation of flux values were adopted following the approach of Sourav S. et al. [9]. These preprocessing steps are essential for preparing the input data for sequential architectures, including Bidirectional Gated Recurrent Unit (GRU), Bidirectional LSTM, and TCN, all of which rely on structured temporal input for effective reconstruction of GRB LC.

### *2.2. Bidirectional Long Short-Term Memory Architecture*

BiLSTM model was selected as a baseline due to its proven effectiveness in modeling complex temporal dependencies, particularly in irregular and noisy time series. BiLSTM architecture is well-suited for capturing both forward and backward temporal contexts, which is especially valuable when reconstructing GRB LC with observational gaps. The BiLSTM network extends the standard LSTM architecture by incorporating two parallel recurrent layers that process the input sequence in both forward and backward directions (Fig. 1). This bidirectional processing enables the model to capture context from both past and future time steps, improving its ability to learn temporal dependencies. The original BiLSTM architecture was introduced by Schuster and Paliwal [33]. Our implementation consists of five stacked BiLSTM layers, each incorporating internal memory mechanisms that allow the network to retain information across long sequences. The first four layers are configured with *return_sequences=True* to preserve the temporal structure throughout the network's depth. This design enables hierarchical feature extraction across multiple time scales. The final BiLSTM layer outputs a fixed-length representation, which is passed to a fully connected dense layer with a Rectified Linear Unit (ReLU) activation function. This layer maps the learned sequence representation to a single flux prediction value. Each BiLSTM layer comprises 100 hidden units, and the model is trained using the Adam optimizer with mean squared error (MSE) as the loss function. To prevent overfitting, early stopping is applied based on validation loss.

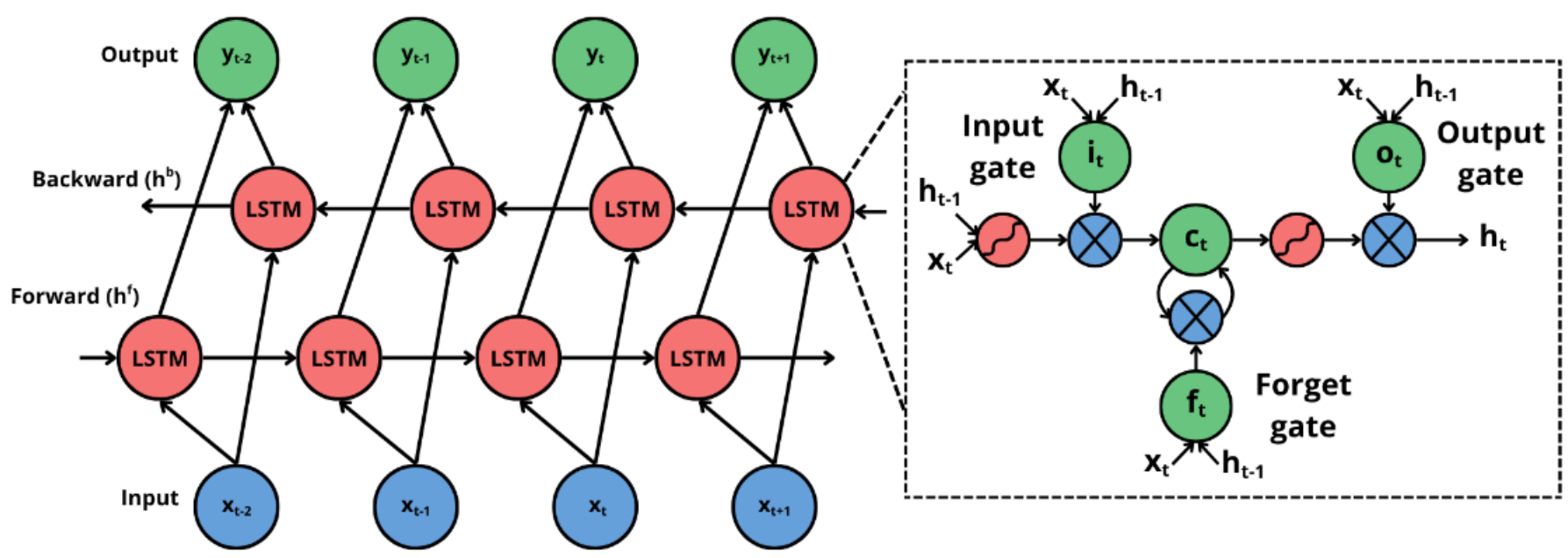

**Fig. 1**. Architecture of a BiLSTM network and the internal structure of an LSTM cell.

Unlike classical regression-based approaches such as the broken power-law (BPL) model - which imposes a predefined parametric form - the BiLSTM model provides greater flexibility in capturing the non-linear and non-stationary nature of GRB LC. It is capable of modeling rapid flux variations, plateau phases, and flaring behavior without requiring strict functional assumptions. The left part of the figure illustrates a BiLSTM model that processes the input sequence in both forward ($h^f$) and backward ($h^b$) directions to generate context-aware outputs at each time step. The right part shows the internal mechanism of an LSTM cell, which includes the input, forget, and output gates. These gates control the flow of information based on the current input $x_t$ and the previous hidden state $h_{t-1}$, maintaining a memory cell state $c_t$ that enables long-term dependency learning.

### *2.3. Temporal Convolutional Network Architecture*

TCN architecture was selected as a non-recurrent alternative for sequence modeling, offering several advantages such as parallel computation, stable gradients, and effective long-range dependency capture through dilated convolutions. TCNs are particularly useful in time-series tasks where causal relationships must be preserved, and memory-efficient modeling is desirable. The TCN framework evaluated by Bai et al. [33] demonstrated strong performance in sequence modeling tasks compared to both LSTM and GRU architectures.

Our implementation includes three 1D convolutional layers with causal padding (Fig. 2) to ensure that predictions at time $t$ do not depend on future inputs. Each layer uses 64 filters and a kernel size of 5, with ReLU activation applied after each convolution. To enable the model to capture longer temporal contexts without increasing kernel size, we employ increasing dilation rates of 1, 2, and 4 respectively across the three convolutional layers. This expands the receptive field, allowing the model to access information from broader time ranges. Following the convolutional layers, the output is flattened and passed to a fully connected dense layer, which outputs a single flux prediction per input. As with the BiLSTM model, the TCN is trained using the Adam optimizer and MSE as the loss function. Early stopping is employed to prevent overfitting, based on validation loss monitoring.

Unlike recurrent models, TCNs do not rely on internal memory states, which makes them less sensitive to vanishing gradients and more efficient for parallelization during training. Furthermore, their ability to handle sequences of varying lengths without explicit unrolling makes them attractive for modeling astrophysical signals with nonuniform sampling. In the context of GRB LC reconstruction, TCNs offer a unique perspective: they capture local and global temporal patterns via stacked convolutional filters while preserving causal dependencies. Their structural simplicity and training efficiency position them as a promising alternative to recurrent networks, particularly when reconstruction speed and computational cost are critical considerations.

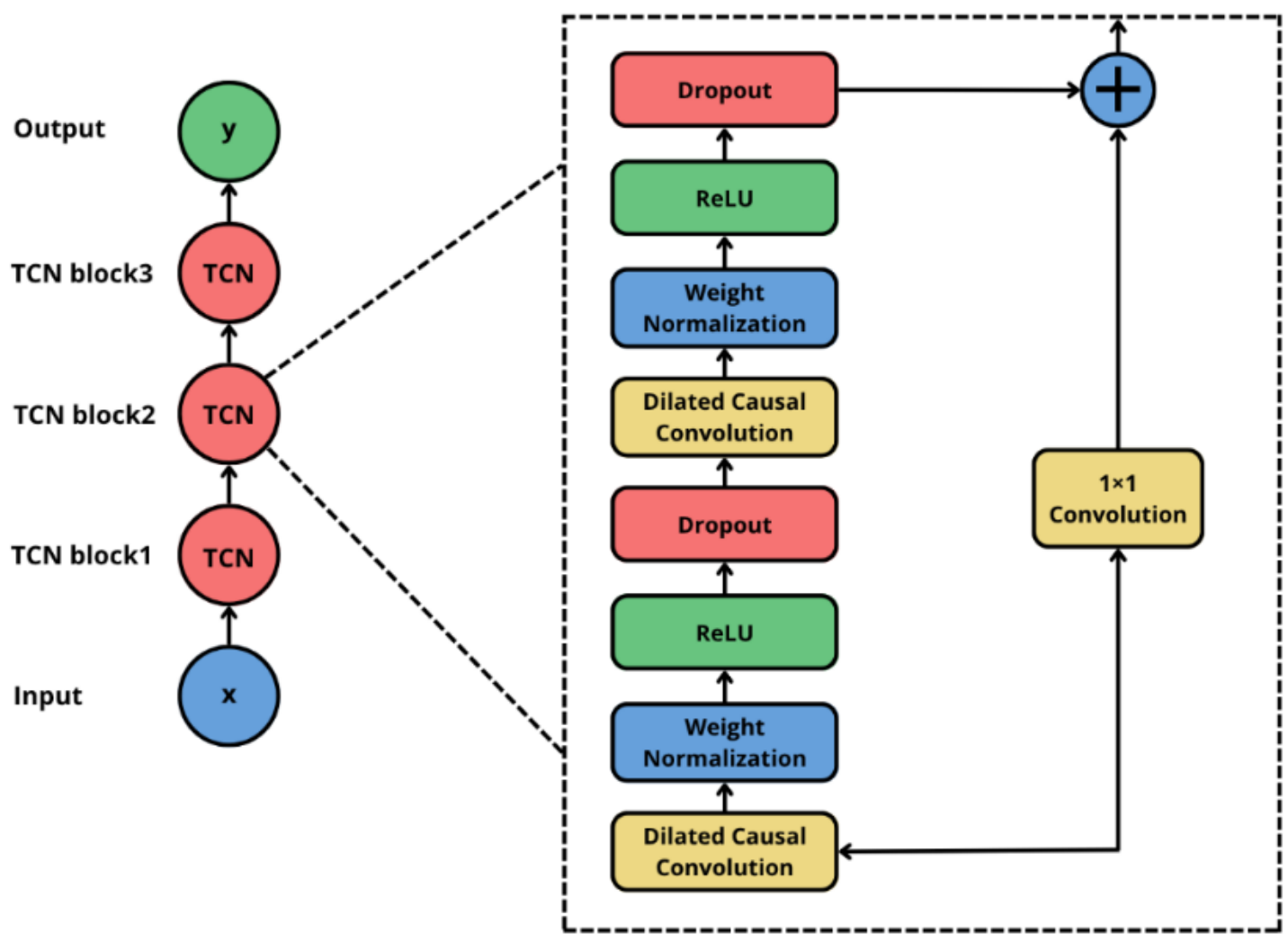

**Fig. 2.** Architecture of a TCN and internal structure of a residual block.

The left part of the figure illustrates a stacked TCN model composed of three residual blocks, which process the input sequence $x$ to produce the output $y$. The right part shows the internal structure of a single

.

residual block, featuring two layers of dilated causal convolutions, each followed by weight normalization, ReLU activation, and dropout. A 1×1 convolution is applied to the input to ensure dimensional compatibility before the residual connection is added. This design enables efficient modeling of long-range dependencies in sequential data.

### *2.4. Bidirectional Gated Recurrent Unit Architecture*

GRU is a simplified recurrent neural architecture proposed by Cho et al. [34] as an alternative to the LSTM. GRUs simplify the memory gating mechanism of LSTMs while maintaining the ability to capture long-term dependencies in temporal data. When combined with a bidirectional architecture, GRUs become particularly effective at modeling sequences where contextual information from both past and future is relevant - a key characteristic of GRB LC with gaps and irregularities.

The BiGRU architecture used in this study consists of two stacked bidirectional GRU layers (Fig. 3). The first layer is configured with *return_sequences=True* to allow the second GRU layer to process temporal dynamics across the full sequence. Each GRU layer contains 64 hidden units, and both layers are wrapped in Keras's Bidirectional wrapper to process input sequences in forward and backward directions. The output of the second GRU layer is fed into a dense layer with a linear activation to produce the final flux prediction.

The model is trained using the Adam optimizer and MSE loss function. An early stopping mechanism is applied based on validation loss to ensure optimal convergence and to avoid overfitting. Due to its simplified gating and reduced computational cost, BiGRU offers faster training and lower memory consumption compared to BiLSTM. This makes it well-suited for applications where computational resources are constrained or where rapid prototyping is needed. Despite its lower complexity, the BiGRU model retains the ability to model complex temporal dependencies and non-linear relationships in the data.

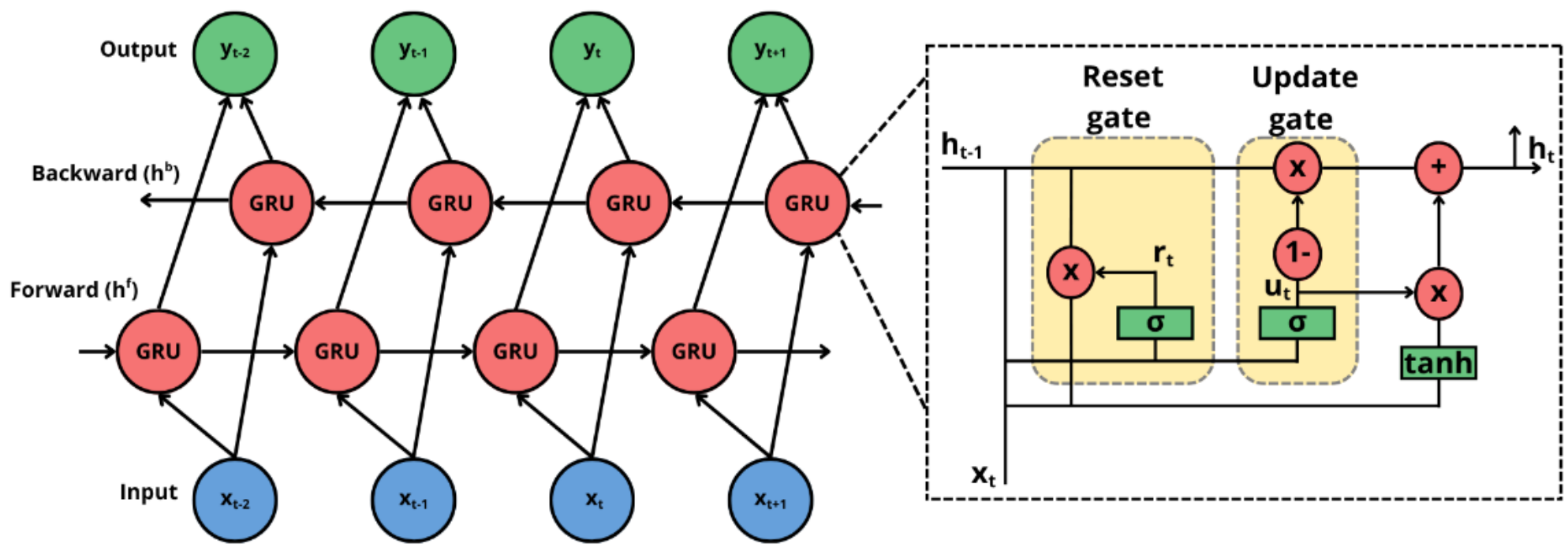

**Fig. 3.** Architecture of a BiGRU network and internal structure of a GRU cell.

The left part of the figure illustrates a BiGRU model that processes the input sequence $x_t$ in both forward ($h^f$ ) and backward ($h^b$) directions to generate context-aware outputs $y_t$ at each time step. The right part shows the internal mechanism of a GRU cell, consisting of the reset gate ($r_t$) and update gate ($u^t$ ), which regulate the flow of information. These gates control the balance between retaining previous memory and incorporating new input, with fewer parameters than LSTM while maintaining similar performance.

### *2.5. Training Procedure*

All models in this study were trained individually on each GRB LC using the preprocessed, interpolated, and upsampled sequences described in Section 2.1. The training process was conducted using the TensorFlow/Keras framework, with GPU acceleration enabled where available. For each model (BiLSTM, BiGRU, and TCN), training was performed using the Adam optimizer and the MSE loss function. To prevent overfitting and improve generalization, we employed early stopping with a patience of five epochs, monitoring validation loss. The data was split into 70% for training and 30% for validation, ensuring that models were evaluated on unseen portions of the LC. During training, each GRB LC was divided into fixed-size batches. The batch size was user-defined and experimentally set to values such as 900 or greater, depending on the number of interpolated points. Each batch was treated as a separate training instance, and the model was trained iteratively across all batches. This approach allowed the models to focus on local temporal patterns while maintaining the ability to generalize across longer sequences.

Inputs were structured in the standard three-dimensional format [samples, timesteps, features], suitable for sequence models. For recurrent models (BiLSTM and BiGRU), the time dimension was explicitly preserved across layers with *return_sequences=True* in intermediate layers. For the TCN model, causal convolutions were used to respect temporal order and avoid leakage of future information into past predictions.

Each model was trained for up to 100 epochs with a batch size of 15, although early stopping typically resulted in faster convergence. After training, the model generated predicted flux values for each input time step. These predictions were then transformed back from logarithmic space into physical flux values for evaluation. The quality of reconstruction was assessed using three standard regression metrics:

Mean Absolute Error (MAE):

$$MAE = \frac{1}{n}\sum_{i=1}^{n} |y_i - \hat{y}_i| , \quad (2)$$

Root Mean Square Error (RMSE):

$$RMSE = \sqrt{\frac{1}{n}\sum_{i=1}^{n}(y_i - \hat{y}_i)^2} , \quad (3)$$

Coefficient of Determination ($R^2$):

$$R^2 = 1 - \frac{\sum_{i=1}^{n}(y_i - \hat{y}_i)^2}{\sum_{i=1}^{n}(y_i - \bar{y}_i)^2} , \quad (4)$$

where n is the total number of samples, i is the index of each sample, $y_i$ is the true (observed) value, $\hat{y}_i$ is the predicted value, and $\bar{y}_i$ is the mean of the true values.

These metrics were calculated between the original and predicted flux values (after inverse transformation), allowing for quantitative comparison between the three model types. In addition, predicted LC were plotted alongside the original observations in log-log scale to visually assess the reconstruction quality. This training pipeline was applied identically to all models, ensuring a fair and controlled comparison across architectural types.

## 3. Results

This section presents the reconstruction outcomes of GRB LC using three deep learning architectures: BiLSTM, TCN, and BiGRU. The analysis focuses on two distinct categories of GRBs: Good GRBs and Break Bump/Bump Flare GRBs, evaluated using both quantitative metrics and visual inspection of the reconstructed LC.

### *3.1 Quantitative Evaluation*

To quantify reconstruction accuracy, three standard regression metrics were employed: MAE, RMSE, and the $R^2$. Tables A1 and A2 summarize the comparative performance of each model across representative bursts from both categories. For Good GRBs, the BiGRU model consistently achieved the lowest MAE and RMSE, along with the highest $R^2$ values, indicating superior reconstruction fidelity. For instance, in GRB050318, BiGRU reduced the RMSE to $1.82 \times 10^{-12}$, compared to $2.49 \times 10^{-12}$ for BiLSTM. Similarly, in GRB060421, all models exhibited strong performance ($R^2 > 0.999$), yet BiGRU maintained a slight edge in RMSE reduction.

In the case of Break Bump and Bump Flare GRBs, BiGRU again demonstrated the highest accuracy, particularly for bursts with complex temporal evolution, such as GRB141221A and GRB140304A. Here, BiGRU's RMSE was more than twofold lower than that of BiLSTM and TCN, while maintaining $R^2 > 0.99$ in most instances.

These results suggest that BiGRU excels in modeling both gradual and abrupt variations in GRB LC. While TCN demonstrated intermediate performance, generally outperforming BiLSTM but falling short of BiGRU, BiLSTM was the least effective, especially for bursts with non-monotonic features, resulting in higher errors and lower $R^2$ values.

.

### *3.2 Visual Evaluation*

Complementing the quantitative analysis, visual inspection of the reconstructed LC (Figures 4-7) provides further insights into model performance. All reconstructions are plotted in logarithmic scale to emphasize dynamic range and temporal structure. Across all examples, BiGRU most accurately reproduces the observed light curves, including plateaus, flares, and decay phases, with minimal deviation. Notably, in GRB050607 and GRB140304A, BiGRU successfully captures complex features (e.g., sharp breaks, flaring activity) with high precision.

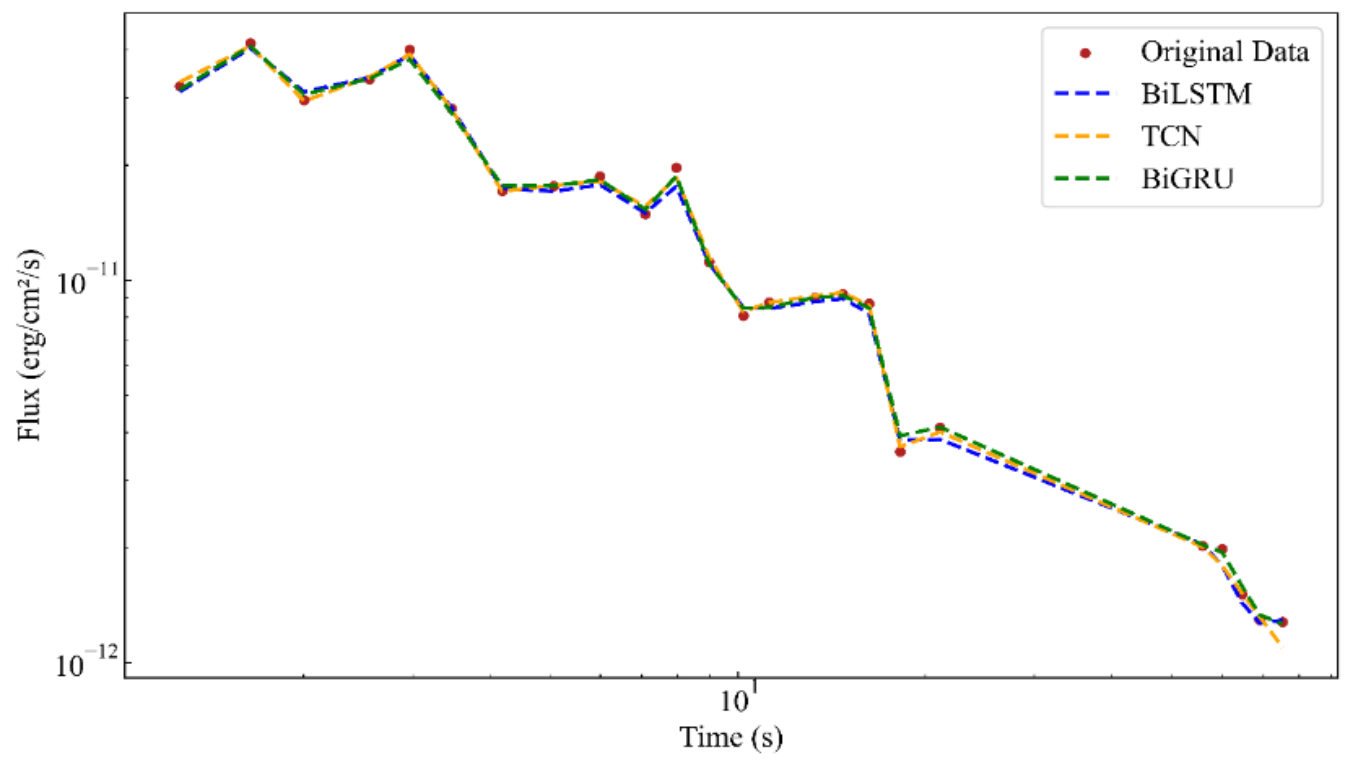

**Fig. 4.** Reconstructed LC of GRB090426 using BiLSTM, TCN, and BiGRU.

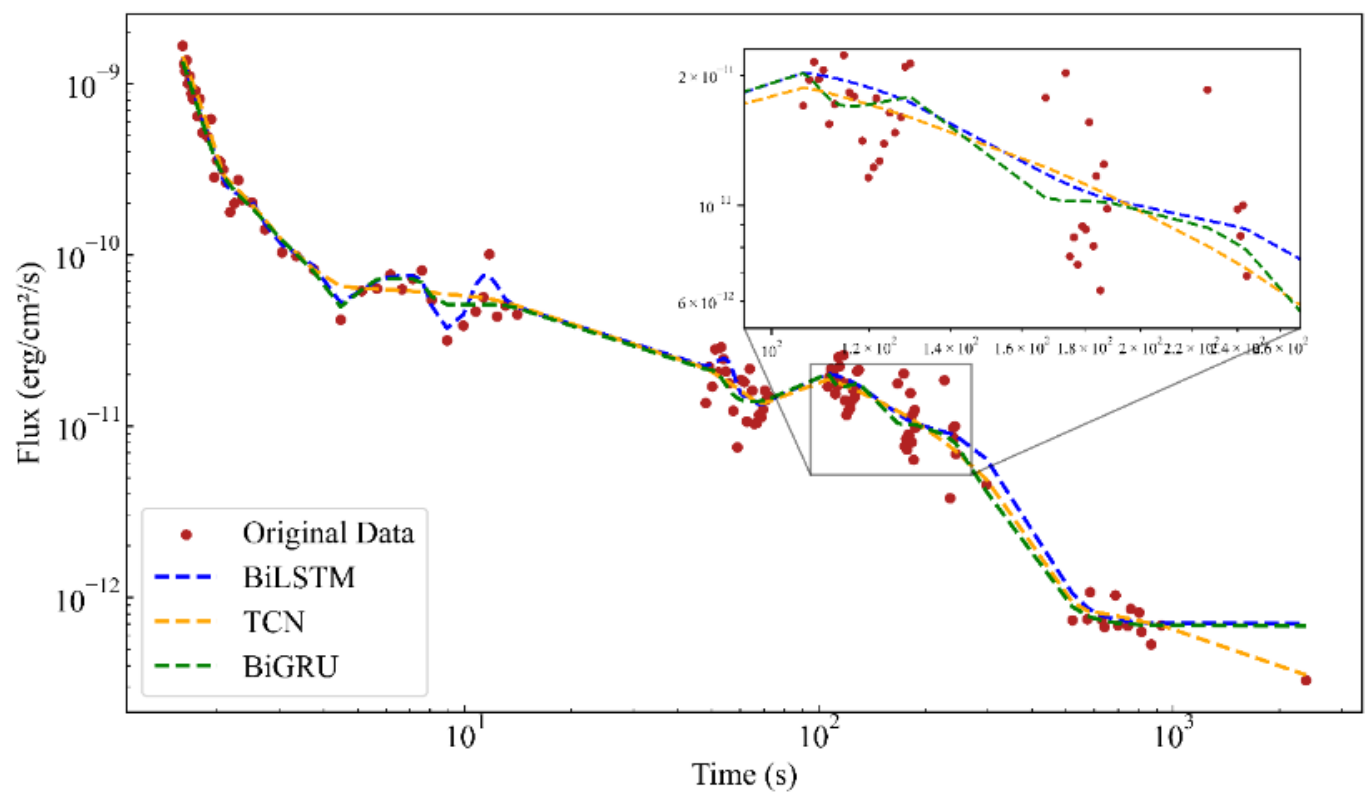

**Fig. 5.** Reconstructed LC of GRB050803 using BiLSTM, TCN, and BiGRU.

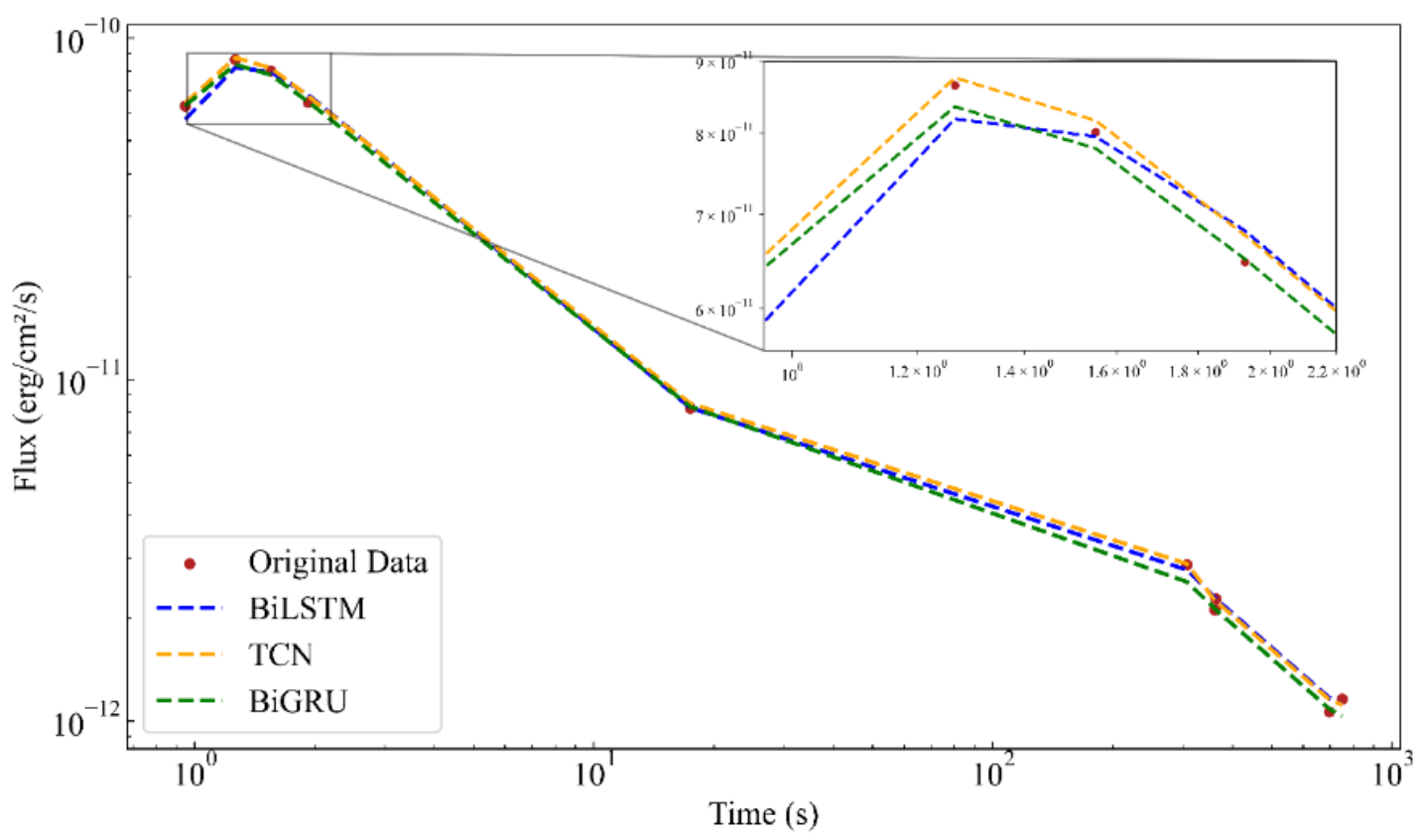

**Fig. 6.** Reconstructed LC of GRB060206 using BiLSTM, TCN, and BiGRU.

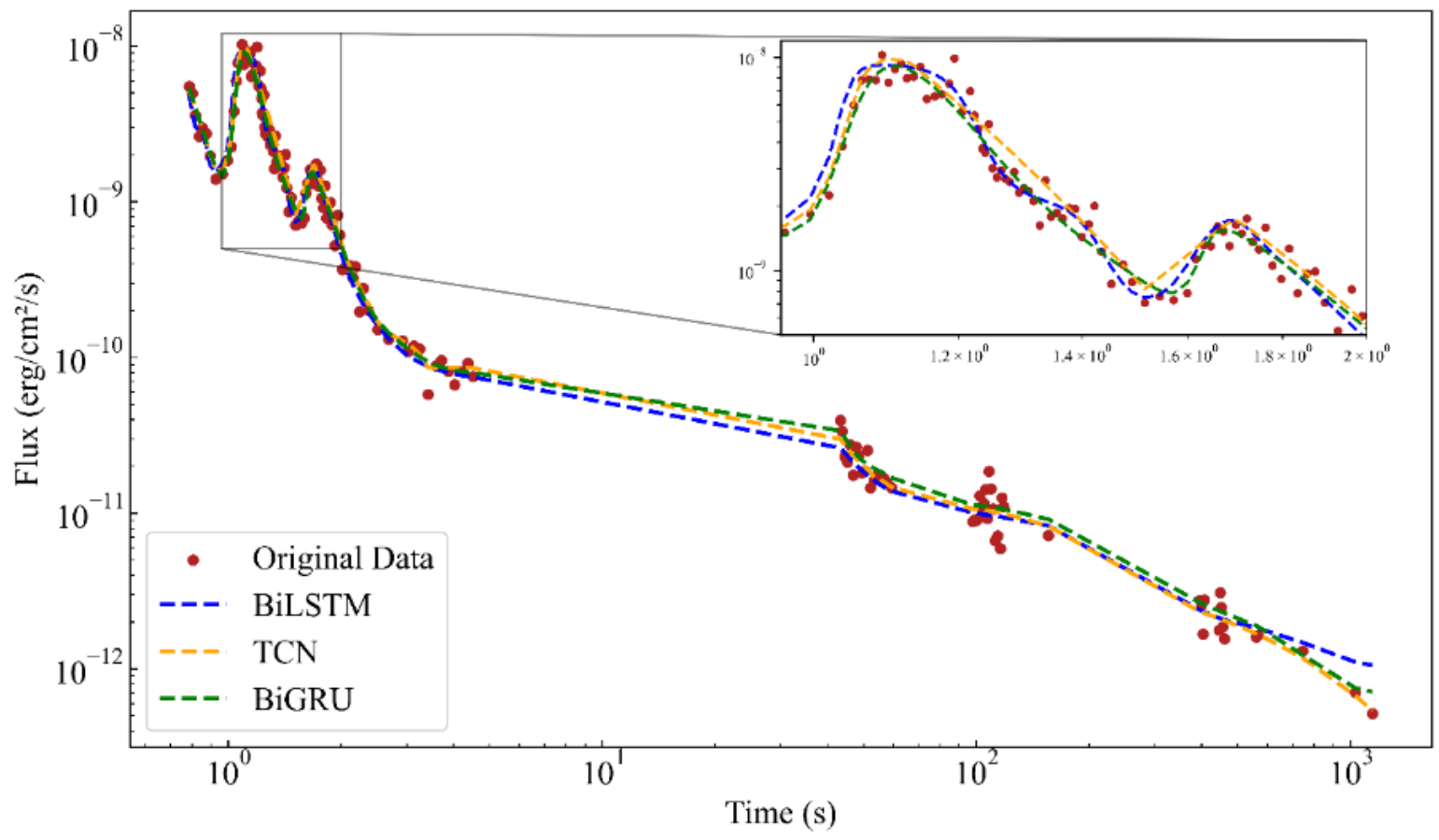

**Fig. 7.** Reconstructed LC of GRB050713A using BiLSTM, TCN, and BiGRU.

TCN generates smoother reconstructions, performing well for bursts with gradual variability but over-smoothing abrupt transitions, particularly in post-break or flaring regions. Conversely, BiLSTM tends to oversimplify variability, often missing localized structures. For example, in GRB141221A, BiLSTM fails to resolve the break feature entirely, whereas BiGRU reconstructs it accurately.

## 4. Conclusion

This study presents a comprehensive evaluation of three deep learning architectures - BiLSTM, BiGRU, and TCN - for reconstructing gamma-ray burst light curves from Swift-XRT X-ray afterglow data. Through systematic preprocessing, including logarithmic normalization, adaptive interpolation, and sequence batching, we established a robust framework for handling observational gaps and noise.

Quantitative analysis using MAE, RMSE, and $R^2$ metrics demonstrated BiGRU's superior performance, achieving up to 27% lower RMSE than BiLSTM and 15% improvement over TCN for complex Break Bump GRBs like GRB141221A, while maintaining MAE $< 2.1 \times 10^{-12}$ erg cm$^{-2}$ s$^{-1}$ and $R^2 > 0.99$ across all GRB types. The BiGRU architecture excelled particularly in capturing both gradual plateau phases and abrupt features such as flares and breaks, as evidenced in challenging cases like GRB140304A. While TCN performed adequately for monotonic decays, it showed systematic over-smoothing of rapid transitions ($\Delta$RMSE $\approx$ 18% for flare-dominated segments), and BiLSTM struggled with fine-scale variability in high-noise regimes. These results position BiGRU as a powerful tool for advancing GRB studies, offering improved afterglow characterization, more accurate luminosity estimations for cosmological applications, and automated identification of physically significant features.

Future research directions should explore hybrid architectures combining BiGRU's temporal modeling with TCN's feature extraction, implement Bayesian uncertainty quantification, and investigate integration of physical constraints through differentiable hydrodynamic priors.

**Conflict of interest statement**

The authors declare that they have no conflict of interest in relation to this research, whether financial, personal, authorship or otherwise, that could affect the research and its results presented in this paper.

**CRediT author statement**

**Zhunuskanov A., Akhmetali A.**: Methodology, Data curation, Conceptualization; **Sakan A.:** Writing–original draft, Visualization; **Ussipov N., Zaidyn M.:** Writing – review & editing, Validation, Supervision; **Zhunuskanov A., Akhmetali A.:** Formal analysis; Investigation. The final manuscript was read and approved by all authors.

**Funding**

This research has been funded by the Science Committee of the Ministry of Science and Higher Education of the Republic of Kazakhstan (Grant AP19674715).

.

**Acknowledgments**

We would like to express our sincerest gratitude to the Department of Electronics and Astrophysics of the Al-Farabi Kazakh National University for supporting this work by providing computing resources (Faculty of Physics and Technology).

## AUTHORS' INFORMATION

**Zhunuskanov Alisher** – Master (Sci.), Department of Physics and Technology, Al-Farabi Kazakh National University, Almaty, Kazakhstan; https://orcid.org/0009-0002-5435-9740; zhunuskanov_alisher1@live.kaznu.kz

**Sakan Aknur** – Master (Sci.), Department of Physics and Technology, Al-Farabi Kazakh National University, Almaty, Kazakhstan; Scopus Author ID: 59194448000; https://orcid.org/0009-0001-8784-4470; aknursakan47@gmail.com

**Akhmetali Almat** – Master (Sci.), Department of Physics and Technology, Al-Farabi Kazakh National University, Almaty, Kazakhstan, Scopus Author ID: 58759186800, https://orcid.org/0009-0005-7254-524X; akhmetalialmat@gmail.com

**Zaidyn Marat** – Bachelor (Sci.), Department of Physics and Technology, Al-Farabi Kazakh National University, Almaty, Kazakhstan; Scopus Author ID: 59194267200; https://orcid.org/0009-0006-8505-7277; zaidyn_marat@live.kaznu.kz

**Ussipov Nurzhan –** PhD, Acting Associate Professor, Department of Physics and Technology, Al-Farabi Kazakh National University, Almaty, Kazakhstan; Scopus Author ID:57226319348, https://orcid.org/0000-0002-2512-3280; ussipov.nurzhan@kaznu.kz

.

**APPENDIX 1**

**Table A1.** Quantitative performance comparison for Good GRBs using BiLSTM, TCN, and BiGRU.

| **GRB ID** | **Model** | **MAE** | **RMSE** | **$R^2$** |
|---|---|---|---|---|
| GRB050318 | BiLSTM | $1.46 \times 10^{-12} \pm 0.16 \times 10^{-12}$ | $2.49 \times 10^{-12} \pm 0.28 \times 10^{-12}$ | 0.943 ± 0.012 |
| | TCN | $1.35 \times 10^{-12} \pm 0.09 \times 10^{-12}$ | $2.26 \times 10^{-12} \pm 0.08 \times 10^{-12}$ | 0.955 ± 0.003 |
| | GRU | $1.10 \times 10^{-12} \pm 0.03 \times 10^{-12}$ | $1.82 \times 10^{-12} \pm 0.02 \times 10^{-12}$ | 0.968 ± 0.0008 |
| GRB050607 | BiLSTM | $1.68 \times 10^{-11} \pm 5.52 \times 10^{-12}$ | $3.92 \times 10^{-11} \pm 1.21 \times 10^{-11}$ | 0.9804 ± 0.0127 |
| | TCN | $1.50 \times 10^{-11} \pm 3.59 \times 10^{-12}$ | $3.07 \times 10^{-11} \pm 8.53 \times 10^{-12}$ | 0.9883 ± 0.0060 |
| | GRU | $1.10 \times 10^{-11} \pm 3.66 \times 10^{-12}$ | $2.64 \times 10^{-11} \pm 8.26 \times 10^{-12}$ | 0.9907 ± 0.0050 |
| GRB050713A | BiLSTM | $2.80 \times 10^{-10} \pm 8.43 \times 10^{-11}$ | $6.04 \times 10^{-10} \pm 1.65 \times 10^{-10}$ | 0.9383 ± 0.0319 |
| | TCN | $2.54 \times 10^{-10} \pm 2.16 \times 10^{-11}$ | $5.45 \times 10^{-10} \pm 5.83 \times 10^{-11}$ | 0.9522 ± 0.0117 |
| | GRU | $1.60 \times 10^{-10} \pm 3.87 \times 10^{-11}$ | $3.29 \times 10^{-10} \pm 5.84 \times 10^{-11}$ | 0.9818 ± 0.0060 |
| GRB050915B | BiLSTM | $7.63 \times 10^{-11} \pm 2.39 \times 10^{-12}$ | $1.58 \times 10^{-10} \pm 1.13 \times 10^{-11}$ | 0.9124 ± 0.0126 |
| | TCN | $7.10 \times 10^{-11} \pm 6.63 \times 10^{-12}$ | $1.48 \times 10^{-10} \pm 1.22 \times 10^{-11}$ | 0.9226 ± 0.0126 |
| | GRU | $6.73 \times 10^{-11} \pm 4.22 \times 10^{-12}$ | $1.34 \times 10^{-10} \pm 5.93 \times 10^{-12}$ | 0.9324 ± 0.0058 |
| GRB060105 | BiLSTM | $1.91 \times 10^{-10} \pm 1.79 \times 10^{-11}$ | $3.09 \times 10^{-10} \pm 3.54 \times 10^{-11}$ | 0.9439 ± 0.0124 |
| | TCN | $1.81 \times 10^{-10} \pm 2.50 \times 10^{-12}$ | $2.93 \times 10^{-10} \pm 5.46 \times 10^{-12}$ | 0.9504 ± 0.0020 |
| | GRU | $1.45 \times 10^{-10} \pm 5.55 \times 10^{-12}$ | $2.36 \times 10^{-10} \pm 6.34 \times 10^{-12}$ | 0.9668 ± 0.0019 |
| GRB060421 | BiLSTM | $3.01 \times 10^{-13} \pm 3.41 \times 10^{-14}$ | $4.97 \times 10^{-13} \pm 6.40 \times 10^{-14}$ | 0.9998 ± 0.0001 |
| | TCN | $1.95 \times 10^{-13} \pm 1.04 \times 10^{-14}$ | $2.72 \times 10^{-13} \pm 9.85 \times 10^{-15}$ | 0.9999 ± 0.0000 |
| | GRU | $1.96 \times 10^{-13} \pm 2.77 \times 10^{-14}$ | $3.71 \times 10^{-13} \pm 5.77 \times 10^{-14}$ | 0.9999 ± 0.0001 |

**APPENDIX 2**

**Table A2.** Quantitative performance comparison for Break Bump / Bump Flare GRBs.

| GRB ID | Model | MAE | RMSE | $R^2$ |
|---|---|---|---|---|
| GRB060206 | BiLSTM | $1.31 \times 10^{-12} \pm 0.48 \times 10^{-12}$ | $2.24 \times 10^{-12} \pm 0.81 \times 10^{-12}$ | 0.996 ± 0.003 |
| | TCN | $1.47 \times 10^{-12} \pm 1.32 \times 10^{-12}$ | $2.20 \times 10^{-12} \pm 2.12 \times 10^{-12}$ | 0.994 ± 0.010 |
| | GRU | $4.16 \times 10^{-13} \pm 1.10 \times 10^{-13}$ | $6.70 \times 10^{-13} \pm 1.55 \times 10^{-13}$ | 0.9996 ± 0.0002 |
| GRB141005A | BiLSTM | $2.28 \times 10^{-12} \pm 9.25 \times 10^{-13}$ | $3.80 \times 10^{-12} \pm 1.35 \times 10^{-12}$ | 0.9895 ± 0.0069 |
| | TCN | $2.54 \times 10^{-12} \pm 6.80 \times 10^{-13}$ | $4.19 \times 10^{-12} \pm 1.24 \times 10^{-12}$ | 0.9878 ± 0.0058 |
| | GRU | $8.40 \times 10^{-13} \pm 1.75 \times 10^{-13}$ | $1.44 \times 10^{-12} \pm 2.45 \times 10^{-13}$ | 0.9985 ± 0.0005 |
| GRB141221A | BiLSTM | $9.04 \times 10^{-12} \pm 3.86 \times 10^{-12}$ | $1.98 \times 10^{-11} \pm 8.08 \times 10^{-12}$ | 0.9896 ± 0.0066 |
| | TCN | $1.52 \times 10^{-11} \pm 3.25 \times 10^{-12}$ | $3.16 \times 10^{-11} \pm 5.80 \times 10^{-12}$ | 0.9767 ± 0.0085 |
| | GRU | $3.15 \times 10^{-12} \pm 1.77 \times 10^{-13}$ | $5.96 \times 10^{-12} \pm 8.47 \times 10^{-13}$ | 0.9992 ± 0.0002 |
| GRB 140713A | BiLSTM | $2.93 \times 10^{-11} \pm 2.55 \times 10^{-12}$ | $5.07 \times 10^{-11} \pm 5.14 \times 10^{-12}$ | 0.9564 ± 0.0088 |
| | TCN | $3.26 \times 10^{-11} \pm 1.04 \times 10^{-12}$ | $5.27 \times 10^{-11} \pm 2.31 \times 10^{-13}$ | 0.9541 ± 0.0015 |
| | GRU | $2.055 \times 10^{-11} \pm 1.78 \times 10^{-12}$ | $3.67 \times 10^{-11} \pm 3.84 \times 10^{-12}$ | 0.9680 ± 0.0123 |
| GRB 050803 | BiLSTM | $1.86 \times 10^{-11} \pm 0.27 \times 10^{-11}$ | $5.33 \times 10^{-11} \pm 0.36 \times 10^{-11}$ | 0.971 ± 0.004 |
| | TCN | $1.87 \times 10^{-11} \pm 0.07 \times 10^{-11}$ | $4.53 \times 10^{-11} \pm 0.15 \times 10^{-11}$ | 0.979 ± 0.001 |
| | GRU | $1.29 \times 10^{-11} \pm 0.14 \times 10^{-11}$ | $3.21 \times 10^{-11} \pm 0.47 \times 10^{-11}$ | 0.988 ± 0.004 |
| GRB 140304A | BiLSTM | $4.05 \times 10^{-11} \pm 9.26 \times 10^{-12}$ | $9.61 \times 10^{-11} \pm 2.35 \times 10^{-11}$ | 0.9469 ± 0.0274 |
| | TCN | $5.28 \times 10^{-11} \pm 4.63 \times 10^{-12}$ | $1.01 \times 10^{-10} \pm 6.51 \times 10^{-12}$ | 0.9458 ± 0.0047 |
| | GRU | $2.82 \times 10^{-11} \pm 4.59 \times 10^{-12}$ | $6.36 \times 10^{-11} \pm 1.00 \times 10^{-11}$ | 0.9704 ± 0.0087 |
| GRB 140713A | BiLSTM | $2.93 \times 10^{-11} \pm 2.55 \times 10^{-12}$ | $5.07 \times 10^{-11} \pm 5.14 \times 10^{-12}$ | 0.9564 ± 0.0088 |
| | TCN | $3.26 \times 10^{-11} \pm 1.04 \times 10^{-12}$ | $5.27 \times 10^{-11} \pm 2.31 \times 10^{-13}$ | 0.9541 ± 0.0015 |
| | GRU | $2.055 \times 10^{-11} \pm 1.78 \times 10^{-12}$ | $3.67 \times 10^{-11} \pm 3.84 \times 10^{-12}$ | 0.9763 ± 0.0049 |
| GRB 090426 | BiLSTM | $5.077 \times 10^{-13} \pm 1.39 \times 10^{-13}$ | $9.09 \times 10^{-13} \pm 3.33 \times 10^{-13}$ | 0.9935 ± 0.0039 |
| | TCN | $4.073 \times 10^{-13} \pm 6.05 \times 10^{-14}$ | $6.72 \times 10^{-13} \pm 7.45 \times 10^{-14}$ | 0.9971 ± 0.0007 |
| | GRU | $2.67 \times 10^{-13} \pm 1.47 \times 10^{-13}$ | $4.16 \times 10^{-13} \pm 2.00 \times 10^{-13}$ | 0.9986 ± 0.0016 |
| GRB140709A | BiLSTM | $4.65 \times 10^{-10} \pm 9.20 \times 10^{-11}$ | $7.83 \times 10^{-10} \pm 1.39 \times 10^{-10}$ | 0.9495 ± 0.0180 |
| | TCN | $4.29 \times 10^{-10} \pm 9.28 \times 10^{-12}$ | $7.33 \times 10^{-10} \pm 2.77 \times 10^{-11}$ | 0.9571 ± 0.0031 |
| | GRU | $3.72 \times 10^{-10} \pm 7.42 \times 10^{-11}$ | $6.14 \times 10^{-10} \pm 1.39 \times 10^{-10}$ | 0.9680 ± 0.0123 |
| GRB050712 | BiLSTM | $2.40 \times 10^{-11} \pm 6.27 \times 10^{-12}$ | $4.22 \times 10^{-11} \pm 1.12 \times 10^{-11}$ | 0.9503 ± 0.0251 |
| | TCN | $2.76 \times 10^{-11} \pm 2.06 \times 10^{-12}$ | $4.95 \times 10^{-11} \pm 5.67 \times 10^{-12}$ | 0.9345 ± 0.0154 |
| | GRU | $1.24 \times 10^{-11} \pm 3.34 \times 10^{-12}$ | $2.24 \times 10^{-11} \pm 6.28 \times 10^{-12}$ | 0.9854 ± 0.0083 |